\input{aipcheck}

\documentclass[
    ,final            
  ]
  {aipproc}

\layoutstyle{8x11double}

\usepackage{axodraw,color}
\usepackage{graphicx}   
\usepackage{subfigure}  
\usepackage[latin1]{inputenc} 
\usepackage{rotfloat} 

\newcommand{\eqa} {\begin{eqnarray} }
\newcommand{\eqe} {\end{eqnarray}}
\newcommand{\beq} {\begin{equation}}
\newcommand{\eeq} {\end{equation}}

\newcommand{\lsp} {\tilde{\chi}_1^0}

\begin{document}

\title{The Passage of Ultrarelativistic Neutralinos through the Matter of the Moon\vspace{-0.3cm}}

\classification{12.60.Jv, 14.80.Ly \vspace{-0.4cm} }

\author{Sascha Bornhauser}{
  address={Department of Physics \& Astronomy, University of New Mexico, Albuquerque, NM 87131, USA}
}



\begin{abstract}
I consider the prospect to use the outer layer of the Moon as a 
detector volume for ultra--high energy (UHE) neutrino fluxes and the flux
of the lightest neutralino which I assume is the lightest supersymmetric particle
(LSP). For this purpose, I calculate the event rates of these fluxes
for top--down scenarios. I show that a suitable experiment for the detection 
of radio waves might be able to detect sufficient event rates after a measurement period of 
one year. 
\end{abstract}

\maketitle

\vspace{-0.9cm}
\section{Introduction}
\vspace{-0.3cm}
A promising idea, which was first suggested by Dagkesamanskii and 
Zheleznyk \cite{Dagkesamanskii}, for the detection of UHE particle fluxes 
like ultrarelativistic neutralinos or neutrinos is the 
measurement of radio waves when these particles hit the Moon  
\cite{gorhamandcoexp}. These 
radio waves are produced due to the Askaryan effect \cite{askara} 
and the emission
of Cerenkov radiation, respectively. UHE particles cause a cascade of 
secondary particles when they are interacting in the Moon's matter. This
cascade develops a cloud of negative charge in a dense dielectric medium
because electrons are entrained from the surrounding matter. As a result,
Cerenkov light is produced since these electrons are moving with a 
velocity which is faster than the velocity of light in the medium. Coherence
builds up for the range of wavelengths which is about the dimension of the
cloud; the wavelengths of radio frequencies are just comparable to 
the dimension of the electron shower \cite{James:2008ff}. 
Therefore we can use a part of 
the outer layer of the Moon as an effective detector volume.
\vspace{-0.5cm}
\section{Equations for the Event Rates}
\vspace{-0.3cm}
This section outlines the derivation of the event rates ${\cal N}$ using the Moon as a detector; I refer to 
\cite{Bornhauser:2006ve,Bornhauser:2007sw,dissertation} for a detailed derivation. 
The event rates for UHE higgsino--like LSPs where the cross section $\sigma_{t}$ is dominated by
the t--channel contributions \cite{Bornhauser:2006ve} is given by:
 \eqa\label{ntch}
{\cal N}_t &=& \int_{E_{\rm min}}^{E_{\rm max}} dE_{\rm vis} 
             \int_{{\bf X}_{\rm min}}^{{\bf X}_{\rm max}} d{\bf X} 
           \int_0^1 dy \frac {1}{y} F_{\lsp} (\frac {E_{\rm vis}} {y}, {\bf X}) \cdot\nonumber\\ 
      &&\left( G_{\lsp}^{\rm NC}(E_{\rm vis},y) + G_{\lsp}^{\rm CC}(E_{\rm
           vis},y)\right) {\mathcal V} \, ,
\eqe
where the constant factor ${\mathcal V}$ is given by
\beq \label{vfactor}
{\mathcal V} \equiv 2\pi V_{\rm eff} \epsilon_{\rm DC} t N_A \rho_w J_D \, 
\eeq
and the convolutions of Eq.~(\ref{ntch}) are given by
\eqa \label{Gs}
G_{\lsp}^{NC,CC}(E_{\rm vis},y) &=&
\int_z^{z_{1,{\rm max}}} \frac{d\!z_1}{z_1}
\frac {d\sigma^{NC,CC}_{t_{\tilde \chi}} (\frac {E_{\rm vis}} {y},z_1)}{dz_1} \cdot\nonumber \\
 &&\frac {1} {\Gamma} 
 \left.\frac {d\Gamma_{\tilde \chi_{\rm out}}(z_1\frac
    {E_{\rm vis}} {y}, z_2 = \frac{z}{z_1})}{dz_2} 
\right|_{z = 1-y} \, . 
\eqe
Here, 
$E_{\rm vis}=E_{\tilde \chi^0_{1,{\rm in}}}-E_{\tilde \chi^0_{1,{\rm out}}}=yE_{\tilde \chi^0_{1,\rm in}}$
denotes the visible Energy, where $E_{\tilde \chi^0_{1,{\rm in}}}$ ($E_{\tilde \chi^0_{1,{\rm out}}}$) is the energy of the LSP before 
(after) the scattering ($\lsp$--nucleon scattering always produces another $\lsp$ in the final state \cite{Bornhauser:2006ve}). {\bf  X} denotes the column depth, measured in g/cm$^2$
and $F_{\lsp}$ the differential neutralino LSP flux.
The $\lsp$ flux with matter leads to fluxes of 
$\tilde \chi_1^\pm$ (charged current (CC) scattering) and $\tilde \chi_2^0$ (neutral current (NC) scattering) \cite{Bornhauser:2006ve}, 
collectively denoted by $\tilde \chi_{\rm out}$;
$z_1 = E_{\tilde \chi_{\rm out}} / E_{\tilde \chi^0_{1,{\rm in}}}$
describes the energy transfer from the incoming lightest neutralino to the
heavier neutralino or chargino, and $z_2 = E_{\tilde \chi^0_{1,{\rm out}}} /
E_{\tilde \chi_{\rm out}}$ describes the energy transfer from this heavier
neutralino or chargino to the lightest neutralino produced in its decay. $z_2$
is chosen such that $z \equiv z_1 z_2 = 1 - y$; 
the missing pieces in Eq.(\ref{Gs}) are the total and differential decay spectrum of the
produced $\tilde \chi_{\rm out}$, see \cite{Bornhauser:2006ve}.
Finally, the constant factor ${\mathcal V}$ is given by
\beq \label{vfactor}
{\mathcal V} \equiv 2\pi V_{\rm eff} \epsilon_{\rm DC} t N_A \rho_w J_D \, .
\eeq
Here, $V_{\rm eff}$ is the water equivalent effective volume,
$\epsilon_{DC}$ is the duty cycle (the fraction of time where the experiment
can observe events), $t$ is the observation time, $N_A = 6.022 \times 10^{23}
\mbox{~g}^{-1}$ is Avogadro's number, $\rho_w = 10^6 \mbox{~g} \mbox{m}^{-3}$
is the density of water, and $J_D = \mid\!d\cos\theta_E/d{\bf X}\!\mid$ is the
Jacobian for the transformation $\cos\theta_E \rightarrow {\bf X}(\cos\theta_E)$, where
$\theta_E$ is the Earth--zenith angle.

The event rates for the three species of neutrinos are given by:
\eqa 
 {\cal N_{\nu_{\mu,\tau}}} &=&  \int d{\mathcal M} \frac {1}{y} F_{\nu_{\mu,\tau}}(\frac {E_{\rm               vis}}{y},{\bf X}(\theta^{\prime},\phi^{\prime},r^{\prime},\phi)) \cdot\nonumber \\ 
           &&\left( \frac {d\sigma_{t_{\nu}}^{\rm NC}(\frac {E_{\rm vis}}{y},y)}
           {dy} + \frac {d\sigma_{t_{\nu}}^{\rm CC}(\frac {E_{\rm vis}}{y},y)}
           {dy} \right) {\mathcal V}^{\prime} \, ,\label{back1mt}\\ 
 {\cal N}_{\nu_{e}} &=&  \int d{\mathcal M} \frac {1}{y} F_{\nu_{e}}(\frac {E_{\rm               vis}}{y},{\bf X}(\theta^{\prime},\phi^{\prime},r^{\prime},\phi))  \cdot\nonumber \\ 
           &&\left(  \frac {d\sigma_{t_{\nu}}^{\rm NC}(\frac {E_{\rm vis}}{y},y)}
           {dy} + N_{\nu_e}^{\rm CC}(E_{\rm vis},y) \right) {\mathcal V}^{\prime} \, ,\label{back1e}
\eqe
where
\eqa
 {\cal N}_{\nu_{e}}^{\rm CC}(E_{\rm vis},y) = \delta\!(y-1)\sigma_{t_{\nu}}^{\rm CC}(E_{\rm vis}) \, ,
\eqe
where ${\mathcal V}$ is now replaced by
\beq \label{mvfactor}
 {\mathcal V} ^{\prime}\equiv \epsilon_{\rm DC_M} t N_A \rho_M  \, .
\eeq
and  $\int d{\mathcal M} $ is a placeholder for the six integrations:
\eqa \label{mback2}
\int d{\mathcal M} &=& \int_{E_{\rm min}}^{E_{\rm max}} dE_{\rm vis} 
           \int_0^1 dy 
            \int_0^{\pi/2}d\theta^{\prime}
            \underbrace{\int_0^{2\pi}d{\phi}^{\prime}}_{=2\pi} \cdot\nonumber \\
            &&\int_{r_{M-10 m}}^{r_M}d{r^{\prime}}{r^{\prime}}^2\sin{\theta}^{\prime}
            \int_0^{2\pi}d\phi
            \int_{53.5^{\circ}}^{56.5^{\circ}}d\theta\sin\theta \nonumber \, .
\eqe
The primed variables 
$\theta^{\prime},\phi^{\prime},r^{\prime}$ denote the spherical coordinates for the 
integration over the allowed volume of the Moon, c.~f.~ \cite{dissertation}; 
the unprimed variables $\phi,\theta$ are used for the 
parametrization of the cone of all 
particle trajectories which can be detected at Earth \cite{dissertation}.
$\theta$ denotes the angle 
of incidence (roughly the Cerenkov light angle $\theta_C$) of the UHE 
particles with respect to the $z-$axis of the spherical coordinates,
which is orientated in direction to the Earth; $\phi$ determines the exact position 
of a trajectory on the circle around the $z-$axis of the cone for a fixed value of $\theta$. 
The density of the surface layer of the Moon (regolith) is given by 
$\rho_M = 1.7\cdot 10^6 \mbox{~g} \mbox{m}^{-3}$ \cite{Stal:2006te,James:2008ff}.
\vspace{-0.7cm}
\section{Numerical Results}
\vspace{-0.3cm}
I assume that an experiment for the detection of radio waves produced by 
Cerenkov radiation can cover one half of the Moon's surface. From this 
I deduce that one has an effective detector volume of about $320$ teratons, 
if the Cerenkov light can leave the regolith up to depth of roughly $10$ m 
\cite{Stal:2006te,James:2008ff}. 
Furthermore, I expect that the Moon appears 40\% of the time above the
radio telescope and I assume a lower bound for the visible energy of 
$10^{10}$ GeV.

I study top-down scenarios with $X$--particle masses of $M_X=10^{12}$ and $M_X=10^{16}$ GeV for
four different primary decay modes, 
where the corresponding 
fluxes at source
were generated with SHdecay \cite{cyrille}.
The event rates are calculated for all three neutrino flavors and 
higgsino--like 
neutralino LSPs. The results are given in 
Tab.~\ref{tabmoon}. 
\begin{table}
\begin{tabular}{|c||c|c|c|c|} 
\hline
Mode & $N_{\mu,\tau}^{\nu}$ & $N_{e}^{\nu}$ & $N_{\rm total}^{\nu}$ & $N_{\lsp}$\\
\hline
\hline
\multicolumn{5}{|c|}{{\bf $E_{\rm vis}\ge 10^{10}$ GeV, $M_X=10^{12}$ GeV  }}\\
\hline
$q\bar q$ & $0.38$ & $1.70$ & $2.46$ & $0.10$\\
$q\tilde q$ & $0.72$ & $2.81$ & $4.25$& $1.10$  \\
$l\tilde l$ & $14.76$ & $35.52$ & $65.04$ & $60.10$ \\
$5\times q\tilde q$ & $1.66$ & $7.97$ & $11.29$& $1.22$ \\
\hline
\multicolumn{5}{|c|}{{\bf  $E_{\rm vis}\ge 10^{10}$ GeV, $M_X=10^{16}$ GeV }}\\
\hline
$q\bar q$ & $1.97$ & $4.44$ & $8.38$ & $0.04$\\
$q\tilde q$ & $1.27$ & $2.98$ & $5.52$ & $0.05$\\
$l\tilde l$ &$1.46$ & $3.17$ & $6.09$& $0.19$ \\
$5\times q\tilde q$ & $1.13$ & $2.71$ & $4.97$  & $0.07$\\
\hline
\end{tabular}
\caption{Event rates for the scenario H2 of \cite{Bornhauser:2006ve}; 
third column shows the sum of the three neutrino fluxes. 
I show results 
for $X$--particle decays into a quark antiquark pair (``$q \bar q$''), into a quark squark pair (``$q \tilde q$''), into a lepton slepton pair 
(``$l \tilde l$''), and into five quarks and five
squarks (``$5 \times q \tilde q$'').  }
\label{tabmoon}
\end{table}
%
The tau and muon neutrino 
fluxes have the same equations for the event rates $N_{\mu,\tau}^{\nu}$ 
due to their
equal behaviour regarding the energy loss of their corresponding leptons 
produced by CC interactions \cite{dissertation}. Similarly, 
the different properties of electrons with respect to their energy 
loss in matter give rise to electron neutrino 
event rates $N_{e}^{\nu}$ being always higher than $N_{\mu}^{\nu}$. 
It is assumed for Eq.~(\ref{back1e}) that 
electron neutrinos give $100\%$ of their energy to the visible energy when 
they undergo a CC interaction. 
I take the same initial spectrum for all three flavors of neutrinos 
since the total neutrino flux impinging on the Earth roughly split 
up to one third per each flavor due to near--maximal neutrino flavor
mixing. In addition, the change of the initial spectra by reason of
their interaction with Moon's matter
is equal for all three neutrino fluxes, c.~f.~\cite{dissertation}.

As shown in \cite{bdhh2}, the expected neutralino LSP flux depends quite strongly
on $M_X$ as well as on the dominant $X$--particle decay mode. Top--down models predict
rather hard spectra, i.e. $E^3$ times the flux increases with energy. Our fixing
of the normalization of the fluxes through
the (proton) flux at $E = 10^{20}$ eV leads to smaller fluxes at $E
< 10^{20}$ eV as $M_X$ is increased. Moreover, if $M_X$ is not far from its
lower bound of $\sim 10^{12}$ GeV, much of the relevant neutralino flux is
produced early in the parton cascade triggered by $X$ decay, which is quite
sensitive to the primary $X$ decay mode. In contrast, if $M_X \gg 10^{12}$
GeV, in the relevant energy range most LSPs originate quite late in the
cascade; in that case the LSP spectrum is largely determined by the dynamics
of the cascade itself, which only depends on Standard Model interactions, and
is not very sensitive to the primary $X$ decay mode(s).
We also see in Tab.~\ref{tabmoon} that in case of the neutrino fluxes 
all four decay modes, independent of the $X$--particle mass, might lead
to an observable signal. The neutralino LSP fluxes yield only detectable
signals for $M_X=10^{12}$ GeV and the last three decay modes, where the 
decay into a lepton plus a slepton is the most favorable one. 
The reason is that this decay mode leads to a rather small number
of protons produced per $X$ decay, or, put differently, to a large ratio of
the LSP and proton fluxes \cite{cyrille}. Since I normalize to the proton
flux, this then leads to a rather large LSP flux.
As
explained above the event rates 
for the higher $X$--particle mass are quite independent from the primary
decay mode, whereas this correlation is still given for the results of
the lower $X$--particle mass. 

\begin{figure}
  \includegraphics[height=.35\textheight,angle= 270]{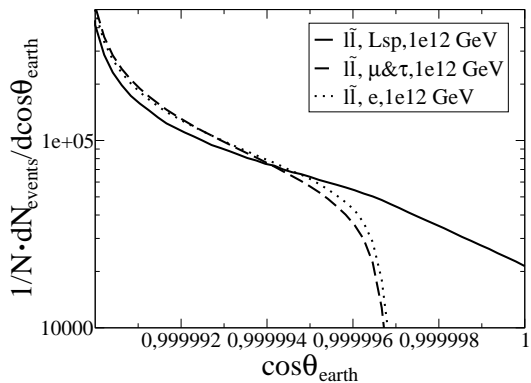}
  \caption{Angular dependence of the 
  signals of neutrinos and higgsino--like neutralino LSPs from
  primary $X \rightarrow l \tilde l$ decays. The solid curve shows the 
 differential event rate for the neutralino flux and the dotted and dashed 
curves show the differential event rate for the electron and tau/muon neutrino
flux, respectively.}
\label{diffEventratecosthetaerde2}
\end{figure}

For $M_X=10^{12}$ GeV and both the second and third decay mode the total event 
rates might have the same order of magnitude with respect to the 
neutralino LSP and neutrino fluxes. This gives rise to the question 
how we can disentangle both signals; this question is 
important for the discrimination between bottom--up and top--down models.
For this purpose, I consider the angular dependence 
of the signals, which is displayed in Fig.~\ref{diffEventratecosthetaerde2}.    
Here, the normalized differential event rate is plotted against the angle 
$\theta_{earth}$ given by the angle relative to the center of the Moon at Earth 
which describes the deviation 
from the connecting line between the Earth's and Moon's center, 
see \cite{dissertation}; 
the smaller 
this angle the smaller is the distance from the connecting line of Earth--Moon.
That means larger averaged travel distances 
for the UHE particles in the Moon's matter. Therefore, the attenuation of the 
neutrino fluxes is higher compared to the neutralino LSP fluxes for such
angles. The former fluxes are negligible for $\theta_{earth}\ge0.999997$
as shown by Fig.~\ref{diffEventratecosthetaerde2}, 
what give rise to the requirement of an angle resolution of at least
$0.14^{\circ}$ for a radio wave antenna experiment if we want to discriminate 
between signals caused by neutralino LSPs and neutrinos. 
\vspace{-0.5cm}
\section{Conclusion}
\vspace{-0.3cm}
I have shown that
a measurement period of one year already might lead 
to detectable event rates of UHE neutrinos for both $X$--particle masses, 
$10^{12}$ and $10^{16}$ GeV, and all four primary 
decay modes.
The above measurement period might lead in the case of UHE neutralino LSPs to
 a measurable signal if the $X$--particles have masses close to their 
lower bound of $\sim 10^{12}$ GeV and decay mainly via the last
three primary decay modes and modes with a large ratio of neutralino 
LSP and proton fluxes, respectively. In case of $X$--particles with 
mass $10^{16}$ GeV one would need at least ten years of detection, even
for the most favorable scenario, to collect a observable event rate. 
The event rates for UHE neutralino LSPs and neutrinos have the same order 
of magnitude for two of the
considered primary decay modes for $X$--particles masses of $10^{12}$ GeV; 
the disentanglement between the neutralino
LSP and neutrino signal is only possible for a radio wave antenna 
experiment having a angle resolution of at least $0.14^{\circ}$. 
\vspace{-0.8cm}
\begin{theacknowledgments}
\vspace{-0.4cm}
 SB thanks the ``Universit\"atsgesellschaft Bonn--Freunde, F\"orderer, 
 Alumni e.V.'' \& the ``Bonn--Cologne Graduate School of Physics and 
 Astronomy'' for financial support.
\end{theacknowledgments}
\vspace{-0.5cm}
\bibliographystyle{aipproc}   


\IfFileExists{\jobname.bbl}{}
 {\typeout{}
  \typeout{******************************************}
  \typeout{** Please run "bibtex \jobname" to optain}
  \typeout{** the bibliography and then re-run LaTeX}
  \typeout{** twice to fix the references!}
  \typeout{******************************************}
  \typeout{}
 }

\end{document}